\documentclass[aps,twocolumn,pra,showpacs,tightenlines]{revtex4}
\usepackage{amssymb}
\usepackage{amsmath}
\usepackage{graphicx}
\usepackage{epsfig}
\usepackage{txfonts}
\usepackage{subfigure}
\usepackage{amsfonts}
\usepackage{CJK}
\usepackage{mathrsfs}
\usepackage{bm}
\makeatletter

\newcommand{\Rmnum}[1]{\expandafter\@slowromancap\romannumeral #1@}
\makeatother

\begin{document}

\begin{CJK*}{GBK}{song}
\title{Controlling quantum coherence of a two-component Bose-Einstein condensate via an impurity atom}
\author{Zhen Li}
\author{Le-Man Kuang\footnote{Author to whom any correspondence should be
addressed. }\footnote{ Email: lmkuang@hunnu.edu.cn}}
\affiliation{Key Laboratory of Low-Dimensional Quantum Structures
and Quantum Control of Ministry of Education,  Department of Physics and SICQEA,
Hunan Normal University, Changsha 410081, China}

\begin{abstract}
We propose a scheme to control  quantum coherence of a two-component Bose-Einstein condensate (BEC) by a single impurity atom immersed in the BEC.   We show that the single impurity atom can act as a single atom valve (SAV) to control  quantum coherence of the two-component BEC. It is demonstrated that the SAV can realize the on-demand control over quantum coherence  at an arbitrary time. Specially, it is found that the SAV can also control higher-order quantum coherence of two-component BEC.
We investigate  the long-time  evolution of quantum coherence of the two-component BEC. It is indicated  that the single impurity atom  can induce collapse and revival phenomenon of quantum coherence of the two-component BEC. Collapse-revival configurations of quantum coherence can be manipulated by the initial-state parameters of the impurity atom and the impurity-BEC interaction strengths.
\end{abstract}

\maketitle \narrowtext
\end{CJK*}

\section{\label{Sec:1}Introduction}

Quantum coherence is of  fundamental and practical significance for the development of quantum physics and quantum technologies, and it has been recognized as  an important physical resource \cite{Herrero,Hu}. Recently, the theory of quantum coherence as a resource has been established by virtue of quantum resource theory to quantify the usefulness of quantum coherence \cite{Baumgratz}. Controlling quantum coherence in many-body systems is highly desirable since quantum coherence  reveals the essence of entanglement and plays a central role in a lot of physical phenomena and applications, such as  quantum  computation \cite{20,21}, quantum metrology \cite{22,23,24,25}, quantum channel discrimination \cite{26,27,28},  quantum phase transitions \cite{31,32}, and quantum thermodynamics \cite{33,34,35,36}.

With the rapid experimental progress in controlling ultracold quantum gases, multicomponent Bose-Einstein condensate (BEC) has now been an active research field in cold atomic physics. Since production of the first binary mixture of condensates consisting of two hyperfine states ($|F,m_f\rangle  = |1,-1\rangle$ and $|2,2\rangle$) of $^{87}$Rb \cite{Myatt}, the two-component BEC has received intensive study in experiments and theories subsequently \cite{Matthews,Hall1,Hall2,Lewandowski,Erhard,Zibold}. A variety of dynamical behaviors have been observed in the two-component BEC \cite{Matthews,Hall1,Hall2,Lewandowski}. Meanwhile, it has been predicted that the two-component Bose gas may exhibit exotic quantum state structures \cite{Ho,Pu,Ao,Cazalilla,Zhou,Ivanov,Tamil} as well as interesting dynamical properties, such as population oscillations between the two states \cite{Matthews,Williams},  quantum self-trapping \cite{Chen}, and quantum correlation effects \cite{Riedel,Gross,Fadel,Laura}.

Recently, much attention has been paid to individual impurities interacting with BEC  due to numerous applications in probing strongly correlated quantum many-body states \cite{Ng,Balewski}, quantum state
engineering \cite{Wang,Schmidt,Heidemann,Mukherjee,Johnson,Yuan1,Song,Lu,Yuan2,Yuan3}, and quantum  metrology \cite{Tan1,Tan2}. In present paper, motivated by the recent progress impurity-doped BEC , we consider a two-level impurity atom which is used to control quantum coherence of the two-component BEC. We demonstrate that the single impurity atom forms a single atom valve to control quantum coherence  of  the two-component BEC. We show that the impurity atom can induce quantum collapse and revival phenomenon for quantum coherence  of  the two-component BEC.

The paper is organized as follows. In Sec. II, we introduce our physical model about single two-level impurity  atom immersed in two-component BEC. In Sec. III, we show a quantum valve effect induced by the impurity atom. We show that the single impurity atom can act as a single atom valve of quantum coherence to control  quantum coherence of the two-component BEC and the valve functionals of the quantum valve are demonstrated in detail.  In Sec. IV, we study the long-time quantum dynamics of quantum coherence for the two-component BEC to reveal impurity-induced collapse and revival  phenomenon for quantum coherence of the two-component BEC. Finally, the concluding section, Sec. V, discusses our main results.

\section{\label{Sec:2} Impurity-doped Two-component BEC model}

In this section we introduce our model system by specifying the Hamiltonian of the impurity-doped two-component BEC system. We consider a two-component BEC in which a single impurity atom is immersed. The  two-component BEC is a binary mixture of condensates
consisting of two hyperfine states, such as $(|F,m_f\rangle=|1,-1\rangle $,
$|2,2\rangle) $ or  $(|F,m_f\rangle=|1,-1\rangle$, $|2, 1\rangle)$  states of  $^{87}$Rb .
The Hamiltonian of the two-component BEC is given by $(\hbar=1)$
\begin{align}
H_{B}=&\int dr^{3}\Psi_{i}^{\dag}(r)\Big[-\frac{1}{2m}\nabla^{2}+V_{i}(r)
+\frac{U_{i}}{2}\Psi_{i}^{\dag}(r)\Psi_{i}(r)\nonumber\\
&+\frac{U_{ij}}{2}\Psi_{j}^{\dag}(r)\Psi_{j}(r)\Big]\Psi_{i}(r),
\end{align}
where $\Psi_{i}(r)$ is the field annihilator operator for the component $i$ at the position $r$, $V_{i}(r)$ is the external trapping potential. Interaction between ultracold atoms are described by a nonlinear self-interaction term $U_{i}=4\pi a_{i}/m$ and a term that corresponds to the nonlinear interaction between different components $U_{ij}=4\pi a_{ij}/m$, where $a_{i}$ is the $s$-wave scattering length of component $i$ and $a_{ij}$ is that between $i$th and $j$th-component condensate, $m$ is the single atom mass in the condensate, $i, j=1,2$.

When  the two-component BEC is confined in a deep potential \cite{57,58}, we can take the single-mode approximation \cite{55,56}   $\Psi_{i}(r)\approx\gamma\varphi_{i}(r)$, where $\gamma=a,b$ and $\varphi_{i}(r)$ are the annihilation operator and the mode function for the component $i$, respectively. Thus, the Hamiltonian of the two-component BEC can be written as
\begin{eqnarray}
H_{B}=\omega_{1}a^{\dag}a+\omega_{2}b^{\dag}b+\chi_{1}(a^{\dag}a)^{2}+\chi_{2}(b^{\dag}b)^{2}+2\chi_{12}a^{\dag}ab^{\dag}b,
\nonumber\\
\end{eqnarray}
where $\omega_{i}$ is the effective harmonic oscillator frequency corresponding to each BEC component in the absence of interaction, the self-interaction strength is $\chi_{1(2)}=\big(2\pi a_{1(2)}/m\big)\int dr^{3}\big|\varphi_{1(2)}(r)\big|^4$, and the inter-component interaction strength is $\chi_{12}=\big(2\pi a_{12}/m\big)\int dr^{3}\big|\varphi^{\dag}_{1}(r)\varphi_{2}(r)\big|^2$.

In our model system the impurity atom is an internal two level system, which we write as an effective spin-1/2. In the following we will also interpret this two-level system as a qubit with two logical states $|0\rangle$ and $|1\rangle$. Cold atom collision physics allows for a situation where the scattering lengths  of the impurity atom and atoms in the BEC are spin-dependent \cite{OMandel}. The Hamiltonian of the single two-level impurity atom  is given by
\begin{eqnarray}
H_{s}=\omega_{0}\sigma_z,
\end{eqnarray}
where the  Pauli operator $\sigma_z$ is defined by $\sigma_z=|1\rangle\langle 1|-|0\rangle\langle 0|$.

The single impurity atom interacts with the two-component BEC via coherent collisions with the following Hamiltonian \cite{Ng,MBruderer}
\begin{eqnarray}
H_{I}=(\sigma_{z}+1)(\lambda_{1}a^{\dag}a+\lambda_{2}b^{\dag}b),
\end{eqnarray}
where the interacting strength is given by $\lambda_{i}=\pi a_{ei}\int dr^{3}|\phi_{e}(r)\varphi_{i}(r)|^{2}/M_{ei}$ with $\phi_{e}(r)$ being the wave function of the impurity atom,   $a_{ei}$   the $s$-wave scattering length between the impurity atom in the upper state $|1\rangle$ and the $i$-th component BEC,  and $M_{ei}=m_{e}m_{i}/(m_{e}+m_{i})$   the reduced mass for impurity atom and BEC atom.

Hence, the total Hamiltonian of the impurity-doped two-component BEC system  can be written as
\begin{align}
H=&\omega_{1}a^{\dag}a+\omega_{2}b^{\dag}b+\omega_{0}\sigma_{z}+\chi_{1}(a^{\dag}a)^{2}+\chi_{2}(b^{\dag}b)^{2}\nonumber\\
&+2\chi_{12}a^{\dag}ab^{\dag}b+(\sigma_{z}+1)(\lambda_{1}a^{\dag}a+\lambda_{2}b^{\dag}b),
\end{align}
where all the interaction strengths can be manipulated by means of a Feshbach-resonance method \cite{60,61,62,63}.

 In what follows, we will study quantum dynamics of quantum coherence of the two-component BEC. We
will show that quantum coherence of the two-component BEC can be controlled through manipulating  initial states of the impurity atom and its interaction with the condensed atoms. Specifically, we will demonstrate that the single impurity atom forms a single atom valve to control quantum coherence  of  the two-component BEC, and find that the impurity atom can induce quantum collapse and revival phenomenon of quantum coherence  in  the two-component BEC system.

\section{\label{Sec:3}Single-atom valve of quantum  coherence for two-component BEC}

In this section we investigate the role of the single impurity atom in controlling quantum coherence  of  the two-component BEC through studying  quantum coherence dynamics of the impurity-doped two-component BEC. We will show that the single impurity atom can act as a single atom valve of quantum coherence to control  quantum coherence of the two-component BEC. We calculate the opening and closing degree (OCD) of the single atom valve (SAV), and show that the SAV can be completely opened and closed by varying initial states of the impurity atom and/or impurity-BEC interactions.

Suppose that the two-component BEC  and the impurity atom  are  initially in a coherent state $|\alpha\rangle\otimes|\beta\rangle$ and a coherent superposition state $\cos\eta|1\rangle+\sin\eta|0\rangle$, respectively. Then the total initial state of the impurity-doped BEC system is given by
\begin{eqnarray}
|\psi(0)\rangle=(\cos\eta|1\rangle+\sin\eta|0\rangle)\otimes|\alpha\rangle\otimes|\beta\rangle,
\end{eqnarray}
where $|\alpha\rangle=D(\alpha)|0\rangle$ and  $|\beta\rangle=D(\beta)|0\rangle$ are two Glauber coherent states for the two-component BEC with the displacement operators being given by  $D(\alpha)=\exp(\alpha a^{\dag}- \alpha^*a)$  and $D(\beta)=\exp(\beta b^{\dag}- \beta^*b)$, respectively.

The wave function of the impurity-doped BEC system  at a time $t$ is given by
\begin{equation}
\vert \psi (t)\rangle =U(t)\vert \psi(0)\rangle,
\end{equation}
where the unitary time evolution operator $U(t)=\exp(-iHt)$ is determined by the Hamiltonian (5).

Making use of Eqs. (5) and (6), it is straightforward to get the total  wave function of the impurity-doped BEC system at a time $t$
\begin{equation}
\vert \psi (t)\rangle= \sin \eta\vert 0\rangle \otimes |\phi_0(\alpha,\beta)\rangle +\cos \eta\vert 1\rangle \otimes |\phi_1(\alpha,\beta)\rangle,
\end{equation}
where we have introduced   generalized coherent states for  boson operators $a$ and $b$ \cite{Titulaer,Bialynicka,Stoler,Kuang1,Kuang2}
\begin{eqnarray}
|\phi_0(\alpha,\beta)\rangle &=& e^{-\frac{\vert \alpha\vert^{2}+\vert\beta \vert ^{2}}{2
}} \sum_{n,m=0}^{\infty }e^{-i\theta_0(n,m)t}\frac{\alpha ^{n}\beta ^{m}}{\sqrt{n!m!}}|n,m\rangle, \nonumber\\
|\phi_1(\alpha,\beta)\rangle &=& e^{-\frac{\vert \alpha\vert^{2}+\vert\beta \vert ^{2}}{2
}} \sum_{n,m=0}^{\infty }e^{-i\theta_1(n,m)t}\frac{\alpha ^{n}\beta ^{m}}{\sqrt{n!m!}}|n,m\rangle,
\end{eqnarray}
where the two running frequency functions are defined by
\begin{eqnarray}
\theta_0(n,m)&=&-\omega _{0}+\omega _{1}n+\omega_{2}m \nonumber\\
&& +\chi _{1}n^{2}+\chi _{2}m^{2}+2\chi _{12}nm, \nonumber\\
\theta_1(n,m)&=&\omega _{0}+(\omega _{1}+2\lambda_{1})n+(\omega_{2}+2\lambda_{2})m \nonumber\\
&& \chi _{1}n^{2}+\chi _{2}m^{2}+2\chi _{12}nm.
\end{eqnarray}

Then, we can obtain the reduced density operator of the two-component BEC by tracing out the impurity part with the following form
\begin{equation}
\rho(t)=\cos ^{2}\eta  |\phi_1(\alpha,\beta)\rangle \langle \phi_1(\alpha,\beta)| +\sin ^{2}\eta  |\phi_0(\alpha,\beta)\rangle \langle \phi_0(\alpha,\beta)|.
\end{equation}

The first-order quantum coherence of the two-component BEC can be characterized by one element of the reduced single-particle density matrix $|\rho_{12}|$ with definition \cite{YXHuang}
\begin{equation}
C_1=\frac{1}{N}|\langle b^{\dag}a\rangle|,
\end{equation}
 where  $N$ is the total number of atoms in the two-component  BEC.

 From Eqs. (11) and (12), we can obtain the mean value $\langle b^{\dag}a\rangle$ at a time $t$
\begin{eqnarray}
\langle b^{\dag}a\rangle&=&\alpha \beta ^{\ast }\Big[ \cos ^{2}\eta e^{-i2( \lambda_{1}-\lambda _{2}) t}+\sin ^{2}\eta \Big]e^{-\vert \alpha\vert ^{2}-\vert \beta \vert ^{2}}\nonumber\\
&&\times e^{\vert \alpha\vert ^{2}e^{-i( 2\chi _{1}-2\chi _{12}) t}}e^{\vert\beta \vert ^{2}e^{-i( -2\chi _{2}+2\chi _{12}) t}}e^{-i(\omega _{1}-\omega _{2}+\chi _{1}-\chi _{2}) t}.
\end{eqnarray}

For the sake of simplicity, we assume that the initial-state parameters $\alpha$ and $\beta$ are real numbers.  Making use of Eqs. (12) and (13) we can get  the analytical expression of  the first-order quantum coherence function
\begin{eqnarray}
C_1(t)&=&\frac{\alpha \beta}{\alpha ^{2}+\beta ^{2}}\sqrt{1-\sin^{2}(2\eta)\sin^{2}(\Delta\lambda t)}   \nonumber\\
&&\times e^{\alpha ^{2}\Big\{\cos\big[2(\chi _{1}-\chi_{12})t\big]-1\Big\}+\beta^{2}\Big\{\cos\big[2(\chi_{2}-\chi_{12})t\big]-1\Big\}},
\end{eqnarray}
where we have introduced the parameter $\Delta\lambda=\lambda_1-\lambda_2$, which denotes the coupling strength difference between impurity and each component condensate. From Eq. (14) we can see that quantum coherence of the two-component BEC  depends on the initial-state parameter $\eta$ of the impurity and the impurity-BEC interaction strengths $\lambda_i$.  Hence, one can control quantum coherence of the two-component BEC  by means of varying
the initial-state parameter of the impurity and the impurity-BEC interaction.

From Eq. (14), we can see that quantum coherence of the two-component BEC can be controlled through changing the initial-state parameter of the impurity atom $\eta$ and the coupling strengths between the impurity atom and condensed atoms. In the following, we will indicate that the  impurity atom can act as a quantum valve to control quantum coherence of the two-component BEC. It is  called as the single atom valve (SAV). This SAV can be used to control the amount of quantum coherence for the two-component BEC at any time in the dynamic evolution of the system under our consideration.

In fact, in the absence of the impurity atom, from Eq. (14) we  can obtain quantum coherence of the two-component BEC as
\begin{eqnarray}
 \tilde{C}_1(t) &=&\frac{\alpha \beta }{\alpha ^{2}+\beta ^{2}}e^{\alpha ^{2}\Big\{\cos\big[2(\chi _{1}-\chi_{12})t\big]-1\Big\}+\beta^{2}\Big\{\cos\big[2(\chi_{2}-\chi_{12})t\big]-1\Big\}}, \nonumber\\
\end{eqnarray}
which indicates that the dynamic evolution of quantum coherence of the two-component BEC is periodical with the period $T=\mathrm{Max}\Big\{\pi/|\chi_1 - \chi_{12}|, \pi/|\chi_2 - \chi_{12}|\Big\}$.

\begin{figure}[htp]
\includegraphics[width=8.5cm,height=6.5cm]{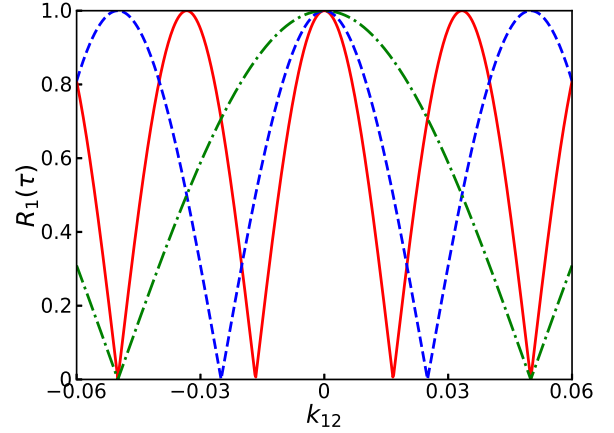}
\caption{(Color online) The OCD of the SAV with respect to the SAV parameter $k_{12}$   when $\eta=\pi/4$. The dot-dashed line, dashed line, and solid line correspond to the scaled time being $\tau=10\pi$, $20\pi$, and $30\pi$, respectively.}\label{fig1}
\end{figure}

In order to measure the controlling degree of quantum coherence, we  can introduce a concept of the opening-closing degree (OCD) of the SAV,  which is defined as the ratio between  quantum coherence of the two-component BEC with the impurity atom and that without the impurity atom
\begin{eqnarray}
R_1(t)=\frac{C_1(t)}{\tilde{C}_1(t)}, \hspace{0.5cm} R_{1}\in [0,1]
\end{eqnarray}
which  gives measurable information about the discharged amount of the first-order quantum coherence of the two-component BEC. The minimum value of the opening-closing  degree  $R_{1}$ is $0$ and the maximum value is $1$, which corresponds to  completely closing (switch-off) and fully opening (switch-on) situation of the SAV, respectively.

Substituting Eqs. (14) and (15) into Eq. (16) we can get
\begin{eqnarray}
R_1(\tau)=\sqrt{1-\sin^{2}(2\eta)\sin^{2}(k_{12}\tau)},
\end{eqnarray}
where we define the coupling strength difference $\Delta\lambda=k_{12}\chi_{12}$ in units of  $\chi_{12}$ and the scaled time $\tau=\chi_{12}t$ with $k_{12}$ and $\tau$ being two dimensionless parameters. Obviously, the initial-state parameter of the impurity atom $\eta$ and the impurity-BEC coupling parameter $k_{12}$ are two characteristic parameters of the SAV.
From Eq. (17) we can see that one can realize on-demand control over quantum coherence of the two-component BEC through adjusting  characteristic parameters of the SAV.
In particular, when $\eta=n\pi$ ($n=0, 1, 2, \cdots$), we have $R_1=1$. This means that  one can obtain the maximal quantum coherence of the two-component BEC at an arbitrary time $\tau$ by adjusting only one characteristic parameter of the SAV, the  initial-state parameter of the impurity atom $\eta$. Hence, the  SAV can be fully opened at an arbitrary time.

\begin{figure}[htp]
\includegraphics[width=9cm,height=7cm]{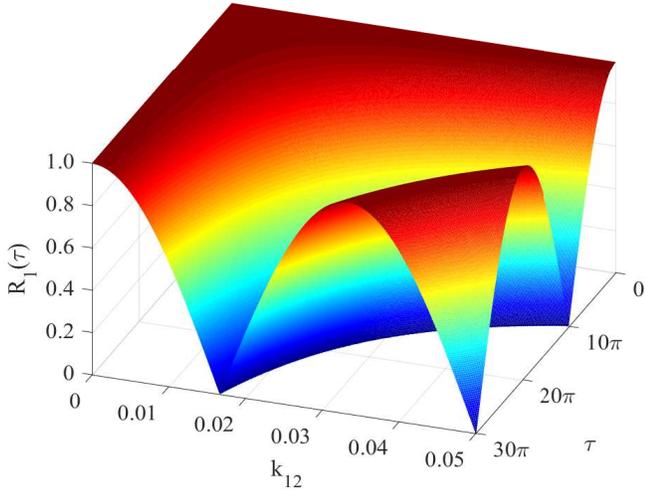}
\caption{(Color online)  The OCD of the SAV with respect to the SAV parameter $k_{12}$  and the scaled time $\tau$ when $\eta=\pi/4$.}\label{fig2}
\end{figure}

We then show how to close the SAV. From Eq. (17) we can see that when $\eta=(2n+1)\pi/4$ ($n=0, 1, 2, \cdots$), and $k_{12}\tau=(2n+1)\pi/2$ ($n=0, 1, 2, \cdots$), we can obtain $R_1=0$. Therefore,  the  SAV can be completely closed at an arbitrary time through changing two characteristic parameters of the SAV  $\eta$ and  $k_{12}$.

In order to show the controllable ability of  the SAV for quantum coherence of the two-component BEC at an arbitrary given time,  in Fig. 1, we have plotted the OCD of the SAV with respect to the SAV parameter $k_{12}$   when $\eta=\pi/4$ for a few particular times of  $\tau=10\pi, 20\pi$, and $30\pi$, respectively. From Fig. 1 we can see that the SAV can continuously work from a completely closed state to a fully opened state  or vice versa by changing the SAV parameter.

In Fig. 2 we have plotted  the influence of the SAV tunable parameter  $k_{12}$ on the controllable functional of the SAV when $\eta=\pi/4$  in the dynamic evolution of the system. From Fig. 2 we can see that it is also possible to realize quantum coherence control of the two-component BEC in the whole regime from $R_{1}=0$ to $R_{1}=1$ through changing the controllable parameter of the SAV $k_{12}$ in the time domain $10\pi\leq \tau \leq 30\pi$.

In Fig. 3 we have plotted  the influence of the SAV tunable parameter  $\eta$ on the controllable functional of the SAV when $k_{12}=0.05$  in the dynamic evolution of the system. From  Fig. 3 we can see that the SAV can be completely closed at the dip points with $R_{1}=0$.

Above analyses indicate that the single impurity atom immersed in the two-component BEC can be used as the SAV of the BEC to control first-order quantum coherence of the two-component BEC efficiently.
One can realize not only switch-on and switch-off of quantum coherence of the two-component BEC, but also on-demand control over the BEC quantum coherence  through adjusting  characteristic parameters of the SAV.

\begin{figure}[htp]
\includegraphics[width=9.0cm,height=7.0cm]{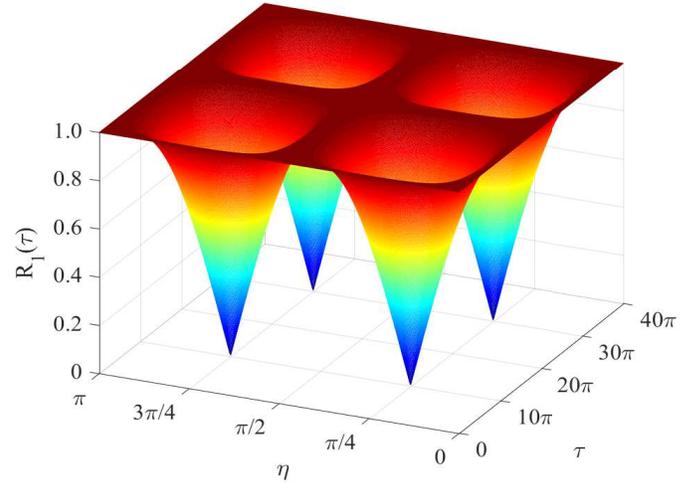}
\caption{(Color online)  The OCD of the SAV with respect to the initial-state parameter of the SAV $\eta$  and the scaled time $\tau$ when the SAV parameter $k_{12}=0.05$.} \label{fig4}
\end{figure}

In what follows, we show that above  SAV of first-order quantum coherence of the two-component BEC is also a SAV of higher-order quantum coherence of the two-component BEC.
The $q$th-order quantum coherence of the two-component BEC can be defined by the following higher-order cross-correlation function  \cite{Opanchuk}
\begin{equation}
C_q=\mathcal{N}\big\vert\langle b^{\dag q}a^{q}\rangle\big\vert, \hspace{0.5cm} (q=1,2,3, \cdots)
\end{equation}
where $\mathcal{N}$ is a normalization factor that ensures the maximum
value of $C_q = 1$ for the optimal case. The normalized ratio gives measurable information about the discharged amount of  $q$th-order quantum coherence of the two-component BEC
\begin{eqnarray}
C_{q}(t)&=&\mathcal{N}(\alpha\beta)^{q}  \sqrt{1-\sin^{2}(2\eta)\sin^{2}(q\Delta\lambda t)}\nonumber\\
& &\times e^{\alpha ^{2}\Big\{\cos\big[2q(\chi _{1}-\chi_{12})t\big]-1\Big\}+\beta^{2}\Big\{\cos\big[2q(\chi_{2}-\chi_{12})t\big]-1\Big\}}.
\end{eqnarray}

In the absence of the impurity atom, from Eq. (19) we can obtain $q$th-order quantum coherence of the two-component BEC as
\begin{equation}
 \tilde{C}_q(t)=\mathcal{N}(\alpha\beta)^{q}
  e^{\alpha ^{2}\Big\{\cos\big[2q(\chi _{1}-\chi_{12})t\big]-1\Big\}+\beta^{2}\Big\{\cos\big[2q(\chi_{2}-\chi_{12})t\big]-1\Big\}}.
\end{equation}

Similar to the case of the first-order quantum coherence, one  can define  the opening-closing degree of the SAV  for the $q$th-order quantum coherence of the two-component BEC as the ratio between the $q$th-order quantum coherence of the two-component BEC with the impurity atom and that without the impurity atom
\begin{eqnarray}
R_q(t)=\frac{C_q(t)}{\tilde{C}_q(t)}, \hspace{0.5cm} R_q\in [0,1].
\end{eqnarray}

Substituting Eqs. (19) and (20) into Eq. (21) and after re-scaling related parameters we can obtain  the OCD of the SAV  for the $q$th-order quantum coherence of the two-component BEC
\begin{eqnarray}
R_q(\tau)=\sqrt{1-\sin^{2}(2\eta)\sin^{2}(qk_{12}\tau)},
\end{eqnarray}
which indicates that the SAV expression (22) about the first-order quantum coherence of the two-component BEC can be recovered when we take $q=1$ in Eq. (22).

Through analyses similar to Eq. (17), from Eq. (22) we can see that the single impurity atom immersed in the two-component BEC can be used as the SAV of the BEC to control higher-order quantum coherence of the two-component BEC efficiently.
One can realize not only switch-on and switch-off function of higher-order quantum coherence of the two-component BEC, but also on-demand control over the BEC higher-order quantum coherence  through adjusting  characteristic parameters of the SAV.

\section{\label{Sec:4} Impurity-induced collapse and revival of quantum coherence}

In this section, we investigate quantum dynamics of quantum coherence of the two-component BEC. We show that the single impurity atom in the two-component BEC can induce  periodic collapse and revival phenomenon of quantum coherence of two-component BEC in the dynamic evolution. This time evolution is a consequence of the quantization of the BEC matter field and the nonlinear interaction between individual atoms. In some sense it can reflect the
quantized structure of the matter wave field and the collisions between individual atoms.

For the sake of simplicity, we re-scale related parameters in terms of $\chi_{12}$ as  $\chi_{1}=k_{1}\chi_{12}$, $\chi_{2}=k_{2}\chi_{12}$, and  $\Delta\lambda=k_{12}\chi_{12}$. Then the expression of first-order quantum coherence of the two-component BEC given by Eq. (14) becomes
\begin{eqnarray}
C_1(\tau)&=&\frac{\alpha \beta} {\alpha ^{2}+\beta ^{2}} \sqrt{1-\sin^{2}(2\eta)\sin^{2}(k_{12}\tau)}\nonumber\\
&&\times e^{\alpha ^{2}[\cos(2k' _{1}\tau)-1]+\beta^{2}[\cos(2k'_{2}\tau)-1]},
\end{eqnarray}
where $\tau=\chi_{12}t$ is the scaled time, $k'_1=k_1-1$ and $k'_2=k_2-1$.

From Eq. (23) we can find that in the absence of the impurity atom  quantum coherence of the two-component BEC has the following simple form
\begin{equation}
\tilde{C_1}(\tau)=\frac{\alpha \beta} {\alpha ^{2}+\beta ^{2}} e^{\alpha ^{2}[\cos(2k' _{1}\tau)-1]+\beta^{2}[\cos(2k'_{2}\tau)-1]},
\end{equation}
which indicates that the dynamic evolution of quantum coherence of the two-component BEC is a simple periodic oscillation with the period of $T=\text{Max}\left(\pi/|k'_1|,\pi/|k'_2|\right)$.

However, in the presence of the impurity atom from Eq. (23) we can see that  the impurity atom can modulate dynamic evolution of quantum coherence of the two-component BEC, and the  whole modulation  factor is given by the OCD $R_1(\tau)=\sqrt{1-\sin^{2}(2\eta)\sin^{2}(k_{12}\tau)}$.  It is the impurity-atom modulation  that breaks down the  simple periodic oscillation of quantum coherence of the two-component BEC and leads to collapse and revival of quantum coherence of the BEC.

\begin{figure}[htp]
\includegraphics[width=8.5cm,height=8.0cm]{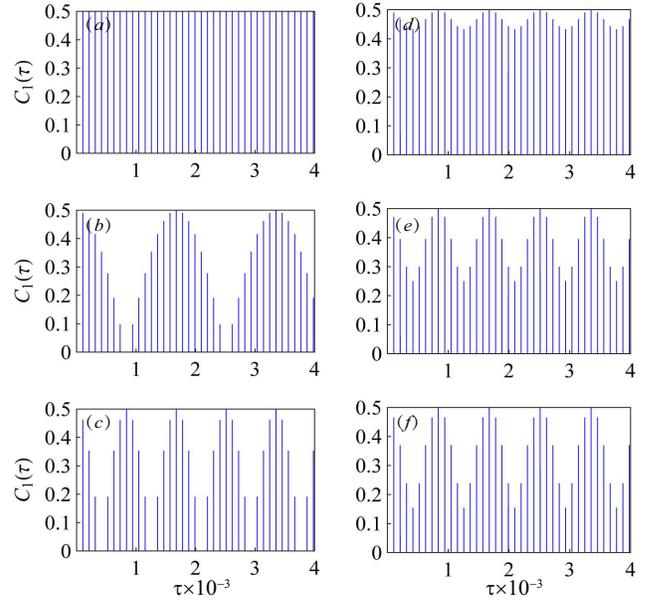}
\caption{(Color online) Quantum coherence of two-component condensates versus scaled time $\tau$ when condensates self-interaction parameters $k_{1}=1.03$, $k_{2}=0.97$ and the atom number of the BEC $\alpha^2=\beta^2=10^5$. When $\eta=\pi/4$, and  $k_{12}=0$, $1.88\times 10^{-3}$, and $3.75\times 10^{-3}$, the quantum coherence evolution of the two-component condensates is given by figs. (a), (b), and (c), respectively. When $k_{12}=3.75\times 10^{-3}$, and  $\eta=\pi/12$, $\pi/6$, $\pi/5$, the quantum coherence evolution of the two-component condensates is given by figs. (d), (e), and (f), respectively.}
\end{figure}

In what follows, we show that quantum coherence of the BEC exhibits collapse and revival phenomenon in the long-time evolution when one of the impurity atom and the BEC is a rapidly changing part and another is a slowly changing part.  Without loss of generality, we consider the situation of  $\alpha=\beta$.  In this case, quantum coherence of the BEC given by Eq. (23) becomes
\begin{equation}
C_1(\tau)=\frac{1} {2} \sqrt{1-\sin^{2}(2\eta)\sin^{2}(k_{12}\tau)} e^{-2\alpha ^{2}[\sin^2(k' _{1}\tau)+\sin^2(k'_{2}\tau)]}.
\end{equation}

 In order to  observe  the impurity atom how to modulate collapse and revival of quantum coherence of the two-component BEC, in Fig. 4 we have plotted the time evolution of quantum coherence of the two-component BEC  when  the BEC is a rapidly changing part while the impurity atom is a slowly changing part with $|k_{12}|\ll (|k'_1|, |k'_2|)$.
From Fig. 4 we can see that collapse and revival phenomenon takes place in the time evolution of quantum coherence of the two-component BEC in the presence of the atom impurity. From Fig. 4(a) we can see that the time evolution of quantum coherence of the two-component BEC is a simple periodic oscillation in the absence of the impurity atom.  Fig. 4(a)-4(c) indicate the time evolution of quantum coherence of the two-component BEC  when $\eta=\pi/4$, $|k'_1|=|k'_2|=0.03$ ($k_1=1.03, k_2=0.97$) and  $\alpha^2=10^5$ for different values of $k_{12}=0, 1.88\times 10^{-3}, 3.75\times 10^{-3}$, respectively. From Fig. 4(a)-4(c) we can see that one can control collapse-revival configuration of quantum coherence of the two-component BEC in the dynamical evolution by varying interaction strength  $k_{12}$. In fact, from Eq. (25) we can see that for the parameter values in Fig. 4(a)-4(c) the envelope curve of the quantum coherence oscillations is determined by the function $\sin^{2}(k_{12}\tau)$. Hence, the collapse and revival time is $\tau_c=\tau_r=\pi/|k_{12}|$.

\begin{figure}[htp]
\includegraphics[width=8.5cm,height=3.5cm]{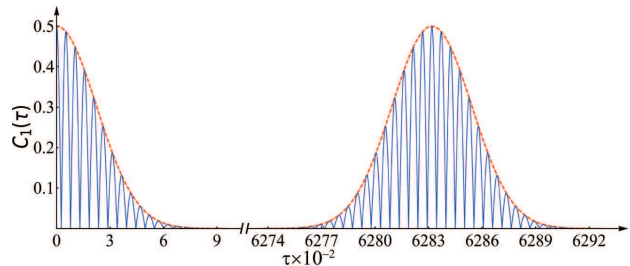}
\caption{(Color online) The long-time evolution of quantum coherence of the two-component BEC when BEC self-interaction parameters  $k'_{1}=-k'_{2}=0.5\times 10^{-5}$, the initial-state parameters   $\alpha^2=\beta^2=10^{5}$, and $\eta=\pi/4$. The dashed line and the solid line correspond to $k_{12}=0$ and $0.06$, respectively.}\label{fig5}
\end{figure}

Fig. 4(d)-4(f) indicate the influence of the initial-state parameter of the impurity atom on the collapse and revival of quantum coherence of the two-component BEC  when $|k'_1|=|k'_2|=0.03$ and  $k_{12}=3.75\times 10^{-3}$ for different values of $\eta=\pi/12$, $\pi/6$, $\pi/5$, respectively. Making use of Eq. (25) from  Fig. 4(d)-4(f)  we can see that the dip value of quantum coherence of the BEC in the collapse regime decreases with the increase of the initial-state parameter $\eta$ in the regime of  $0\leq \eta \leq \pi/4$.

We now consider the situation that  the impurity atom  is a rapidly changing part while the BEC is a slowly changing part. In this case  we have  $ |k_{12}|\gg (|k'_1|, |k'_2|)$ with   $k_{12}$ and $\eta$ being two control parameters.
In Fig. 5 and Fig.6 we have plotted the long-time evolution of quantum coherence of the two-component BEC  when the impurity atom  is a rapidly changing part while the BEC is a slowly changing part with in Fig. 5  the BEC self-interaction parameters  take $k'_{1}=-k'_{2}=0.5\times 10^{-5}$, the initial-state parameters  take $\alpha^2=\beta^2=10^{5}$, and $\eta=\pi/4$. The dashed line and the solid line correspond to $k_{12}=0$ and $0.06$, respectively. In Fig. 6 interaction parameters take $k'_{1}=-k'_{2}=0.5\times 10^{-5}$ and $k_{12}=0.06$, the BEC initial-state parameters take  $\alpha^2=\beta^2=10^{5}$. The dashed line and the solid line correspond to $\eta=0$ and $\pi/6$, respectively. In Fig. 5  the time evolution of  quantum coherence of the two-component BEC without the impurity atom ($k_{12}=0$) is given by the dashed line, which is just the envelope line of quantum coherence of the two-component BEC in the presence of the impurity atom ($k_{12}=0.06$). A similar result is also reflected in Fig. 6.

From Fig. 5  and Fig. 6 we can see that  collapse and revival phenomenon takes place in the long-time evolution of quantum coherence of the two-component BEC. Fig. 5 indicates the dependence of the quantum  collapse and revival on the impurity-BEC interaction $k_{12}$  while Fig.6 reflects the dependence of the quantum  collapse and revival on the initial-state parameter $\eta$.  Fig. 5  and Fig. 6 indicate that    the envelope of the quantum coherence oscillations collapses to nearly zero with the time evolution. However, as time increases  the collapsed quantum coherence can revive till to reaching the maximal coherence. The behavior of collapse and revival of quantum coherence is repeated and the envelope configuration of the oscillations remains unchanged as time increases.

\begin{figure}[htp]
\includegraphics[width=8.5cm,height=3.5cm]{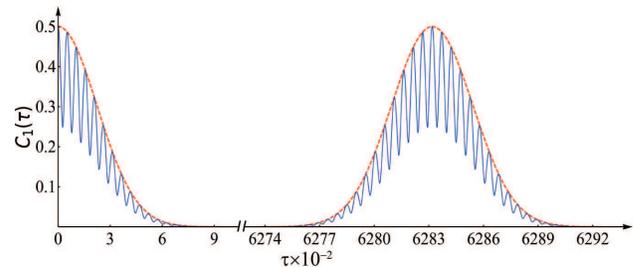}
\caption{(Color online) The long-time evolution of quantum coherence of the two-component BEC when  interaction parameters  $k'_{1}=-k'_{2}=0.5\times 10^{-5}$ and $k_{12}=0.06$, the BEC initial-state parameters   $\alpha^2=\beta^2=10^{5}$. The dashed line and the solid line correspond to $\eta=0$ and $\pi/6$, respectively.}\label{fig6}
\end{figure}

Comparing Fig. 5 with Fig. 6 we can find that the  collapse and revival of quantum coherence controlled by the impurity-BEC interaction $k_{12}$ has the same envelope as that controlled by the initial-state parameter $\eta$. This implies that the $k_{12}$-controlled and $\eta$-controlled  collapse and revival phenomena have the same collapse time and revival time, but they exhibit completely different internal oscillation structures.
It should be pointed that the BEC matter wave field has a quantized structure owing to the granularity of the discrete underlying atoms \cite{Greiner}.   Collapse and revival phenomenon of quantum coherence for the two-component BEC is a manifestation of the quantum nature of the matter-wave field. This point is just like collapse and revival phenomenon of the atomic inversion in the Jaynes-Cummings model for the cavity QED, it reveals the quantum nature of the optical field \cite{Shore,KLM}.
Instead of the simple oscillations of quantum  coherence found in the absence of the impurity atom,  the  collapse and revival of quantum coherence of the two-component BEC takes place only in the presence of the impurity atom.  In this sense the impurity atom can be regarded as a probe to explore  the quantum nature of the BEC matter-wave field.

\section{\label{Sec:6} Summary and conclusion }

We have studied a scheme utilizing a single impurity atom to control quantum coherence of the two-component BEC.  We have shown that the single impurity atom can act as a SAV to control   quantum coherence of the two-component BEC. The SAV can be fully opened and completely closed  by varying initial states of the impurity atom and/or impurity-BEC interactions. In particular, it has been demonstrated that the SAV can realize the on-demand control over quantum coherence of the two-component BEC at an arbitrary time. It has been found that the SAV can control not only  first-order quantum coherence   but also higher-order quantum coherence of two-component BEC.
We have investigated the long-time dynamic evolution of quantum coherence of the two-component BEC. It has been indicated  that the single impurity atom immersed in the two-component BEC can induce  periodic collapse and revival phenomenon of quantum coherence of two-component BEC in the dynamic evolution.  It has been found that collapse-revival configurations of quantum coherence can be manipulated by the initial-state parameters of the impurity atom and  the impurity-BEC interaction strengths. This collapse and revival phenomenon is a consequence of the quantization of the BEC matter field and the nonlinear interaction between individual atoms, it can reflect the quantized structure of the matter wave field and the collisions between individual atoms.  It also may provide information on the long-time entangled dynamics between two components of the BEC.

\acknowledgments
This work is supported by the National Natural
Science Foundation of China under Grants No. 11775075, No. 1143011 and No. 11935006.


\begin{thebibliography}{99}
\bibitem{Herrero} M. Herrero-Collantes, and J. C. Garcia-Escartin, Quantum random number generators, Rev. Mod. Phys. 89, 015004 (2017).

\bibitem{Hu} M. L. Hu, X. Y. Hu, J. C. Wang, Y. Peng, Y. R. Zhang, and H. Fan, Quantum coherence and geometric quantum discord, Phys. Rep. 762,1 (2018).

\bibitem{Baumgratz} T. Baumgratz, M. Cramer, and M. B. Plenio, Quantifying coherence, Phys. Rev. Lett. 113, 140401 (2014).

\bibitem{20} M. Hillery, Coherence as a resource in decision problems: The Deutsch-Jozsa algorithm and a variation, Phys. Rev. A 93, 012111 (2016).

\bibitem{21} D. Deutsch, and R. Jozsa, Rapid solution of problems by quantum computation, Proc. R. Soc. A 439, 553 (1992).

\bibitem{22} S. L. Braunstein, and C. M. Caves, Statistical distance and the geometry of quantum states, Phys. Rev. Lett. 72, 3439 (1994).

\bibitem{23} S. L. Braunstein, C. M. Caves, and G. J. Milburn, Generalized uncertainty relations: theory, examples, and Lorentz invariance, Ann. Phys. 247, 135 (1996).

\bibitem{24} V. Giovannetti, S. Lloyd, and L. Maccone, Quantum-enhanced measurements: beating the standard quantum limit, Science 306, 1330 (2004).

\bibitem{25} V. Giovannetti, S. Lloyd, and L. Maccone, Quantum metrology, Phys. Rev. Lett. 96, 010401 (2006).

\bibitem{26} D. Girolami, T. Tufarelli, and G. Adesso, Characterizing Nonclassical Correlations via Local Quantum Uncertainty, Phys. Rev. Lett. 110, 240402 (2013).

\bibitem{27} A. Farace, A. De Pasquale, L. Rigovacca, and V. Giovannetti, Discriminating strength: a bona fide measure of nonclassical correlations, New J. Phys. 16, 073010 (2014).

\bibitem{28} D. Girolami, Observable Measure of Quantum Coherence in Finite Dimensional Systems, Phys. Rev. Lett. 113, 170401 (2014).



\bibitem{31} G. Karpat, B. \c{C}akmak, and F. F. Fanchini, Quantum coherence and uncertainty in the anisotropic XY chain, Phys. Rev. B 90, 104431 (2014).

\bibitem{32} B. \c{C}akmak, G. Karpat, and F. F.Fanchini, Factorization and criticality in the anisotropic XY chain via correlations, Entropy 17, 790 (2015).

\bibitem{33} F. G. S. L. Brand\~{a}o, M. Horodecki, J. Oppenheim, J. M. Renes, and R.W. Spekkens, Resource theory of quantum states out of thermal equilibrium, Phys. Rev. Lett. 111, 250404 (2013).

\bibitem{34} G.Gour, M. P. M\"{u}ller, V. Narasimhachar, R.W. Spekkens, and N. Y. Halpern, The resource theory of informational nonequilibrium in thermodynamics, Phys. Rep. 583, 1 (2015).

\bibitem{35} J. Goold, M. Huber, A. Riera, L. del Rio, and P. Skrzypczyk, The role of quantum information in
thermodynamics-a topical review, J. Phys. A 49, 143001 (2015).

\bibitem{36} D. Janzing, P. Wocjan, R. Zeier, R. Geiss, and T. Beth, Thermodynamic cost of reliability and low
temperatures: tightening Landauer's principle and the second law, Int. J. Theor. Phys. 39, 2717 (2000).

\bibitem{Myatt}  C. J. Myatt, E. A. Burt, R. W. Ghrist, E. A. Cornell, and C. E. Wieman, Production of two overlapping Bose-Einstein condensates by sympathetic cooling, Phys. Rev. Lett. 78, 586 (1997).

\bibitem{Matthews}  M. R. Matthews, D. S. Hall, D. S. Jin, J. R. Ensher,C. E. Wieman, and E. A. Cornell, F. Dalfovo, C. Minniti, and S. Stringari, Dynamical response of a Bose-Einstein condensate to a discontinuous change in internal state, Phys. Rev. Lett. 81, 243 (1998).

\bibitem{Hall1} D. S. Hall, M. R. Matthews, J. R. Ensher, C. E. Wieman, and E. A. Cornell, Dynamics of component separation in a binary mixture of Bose-Einstein condensates, Phys. Rev. Lett. 81, 1539 (1998).

\bibitem{Hall2}  D. S. Hall, M. R. Matthews, C. E. Wieman, and E. A. Cornell, Measurements of relative phase in two-component Bose-Einstein condensates, Phys. Rev. Lett. 81, 1543 (1998).

\bibitem{Lewandowski}  H. J. Lewandowski, D. M. Harber, D. L. Whitaker, and E. A. Cornell, Observation of anomalous spin-state segregation in a trapped ultracold vapor, Phys. Rev. Lett. 88, 070403 (2002).

\bibitem{Erhard}  M. Erhard, H. Schmaljohann, J. Kronj\"{a}ger, K. Bongs, and K. Sengstock, Measurement of a mixed-spin-channel Feshbach resonance in $\sideset{^{87}}{}{\mathop{\mathrm{Rb}}}$, Phys. Rev. A 69, 032705 (2004).

\bibitem{Zibold}  T. Zibold, E. Nicklas, C. Gross, and M. K. Oberthaler, Classical bifurcation at the transition from Rabi to Josephson dynamics, Phys. Rev. Lett. 105, 204101 (2010).

\bibitem{Ho}  T. L. Ho, and V. B. Shenoy, Binary mixtures of Bose condensates of alkali atoms, Phys. Rev. Lett. 77, 3276 (1996).

\bibitem{Pu}  H. Pu, and N. P. Bigelow, Properties of two-species Bose condensates, Phys. Rev. Lett. 80, 1130 (1998).

\bibitem{Ao}  P. Ao, and S. T. Chui, Binary Bose-Einstein condensate mixtures in weakly and strongly segregated phases, Phys. Rev. A 58, 4836 (1998).

\bibitem{Cazalilla}  M. A. Cazalilla and A. F. Ho, Instabilities in binary mixtures of one-dimensional quantum degenerate gases, Phys. Rev. Lett. 91, 150403 (2003).

\bibitem{Zhou}  L. Zhou, J. Qian, H. Pu, W. P. Zhang, and H. Y. Ling, Phase separation in a two-species atomic Bose-Einstein condensate with an interspecies Feshbach resonance, Phys. Rev. A 78, 053612 (2008).

\bibitem{Ivanov} S. K. Ivanov and A. M. Kamchatnov, Simple waves in a two-component Bose-Einstein condensate, Phys. Rev. E 97, 042208 (2018).

\bibitem{Tamil} R. Tamilthiruvalluvar,  E. Wamba,  S. Subramaniyan, and K. Porsezian, Impact of higher-order nonlinearity on modulational instability in two-component Bose-Einstein condensates, Phys. Rev. E 99, 032202 (2019).

\bibitem{Williams}  J. Williams, R. Walser, J. Cooper, E. Cornell, and M. Holland, Nonlinear Josephson-type oscillations of a driven, two-component Bose-Einstein condensate, Phys. Rev. A 59, R31 (1999).

\bibitem{Chen}  Z. D. Chen, J. Q. Liang, S. Q. Shen, and W. F. Xie, Dynamics and Berry phase of two-species Bose-Einstein condensates, Phys. Rev. A 69, 023611 (2004).



\bibitem{Riedel}   M. F. Riedel, P. B\"{o}hi, Y. Li, T. W. H\"{a}nsch, A. Sinatra, and P. Treutlein, Atom-chip-based generation of entanglement for quantum metrology, Nature 464, 1170 (2010).

\bibitem{Gross}  C. Gross, T. Zibold, E. Nicklas, J. Est\`{e}ve, and  M. K. Oberthaler, Nonlinear atom interferometer surpasses classical precision limit, Nature 464, 1165 (2010).

\bibitem{Fadel}  M. Fadel, T. Zibold, B. D\'{e}camps, P. Treutlein, Spatial entanglement patterns and Einstein-Podolsky-Rosen steering in Bose-Einstein condensates, Science, 360, 409 (2018).

\bibitem{Laura} L. Rosales-Z\'{a}rate,  B. J. Dalton, and M. D. Reid, Einstein-Podolsky-Rosen steering, depth of steering, and planar spin squeezing in two-mode Bose-Einstein condensates, Phys. Rev. A 98, 022120 (2018).

\bibitem{Ng} H. T. Ng, and S. Bose, Single-atom-aided probe of the decoherence of a Bose-Einstein condensate, Phys. Rev. A 78, 023610 (2008).

\bibitem{Balewski} J. B. Balewski, A. T. Krupp, A. Gaj, D. Peter, H. P. B\"{u}chler, R. L\"{o}w, S. Hofferberth, and T. Pfau, Coupling a single electrontoa Bose-Einstein condensate, Nature (London) 502, 664 (2013).

\bibitem{Wang} J. Wang, M. Gacesa, and R. C\^{o}t\'{e}, Rydberg electrons in a Bose-Einstein condensate, Phys. Rev. Lett. 114, 243003 (2015).

\bibitem{Schmidt}  R. Schmidt, H. R. Sadeghpour, and E. Demler, Mesoscopic Rydberg impurity in an atomic quantum gas, Phys. Rev. Lett. 116, 105302 (2016).

\bibitem{Heidemann}  R. Heidemann,  U. Raitzsch, V. Bendkowsky, B. Butscher, R. L\"{o}w, and T. Pfau, Rydberg excitation of Bose-Einstein condensates, Phys. Rev. Lett. 100, 033601 (2008).

\bibitem{Mukherjee}  R. Mukherjee, C. Ates, W. Li, and S. W\"{u}ster,  Phase-imprinting of Bose-Einstein condensates with Rydberg impurities, Phys. Rev. Lett. 115, 040401 (2015).

\bibitem{Johnson} T. H. Johnson,  Y. Yuan,  W. Bao,  S. R. Clark, C. Foot,  and D. Jaksch, Hubbard model for stomic impurities bound by the vortex lattice of a rotating Bose-Einstein condensate, Phys. Rev. Lett. 116, 240402 (2016).

\bibitem{Yuan1} J. B. Yuan, W. J. Lu, Y.J. Song, and L. M. Kuang, Single-impurity-induced Dicke quantum phase transition in a cavity-Bose-Einstein condensate, Sci. Rep. 7, 7404 (2017).

\bibitem{Song}  Y. J. Song, and L. M. Kuang, Controlling decoherence speed limit of a single impurity
atom in a Bose-Einstein-condensate reservoir, Ann. Phys. (Berlin) 531, 1800423 (2019).

\bibitem{Lu} W. J. Lu, Z. Li, and L. M. Kuang, Nonlinear Dicke quantum phase transition and its quantum witness in
a cavity-Bose-Einstein-condensate system, Chin. Phys. Lett. 35, 116401 (2018).


\bibitem{Yuan2}  J. B. Yuan, H. J. Xing, and  L. M. Kuang,  and S. Yi, Quantum non-Markovian reservoirs of atomic condensates engineered via dipolar interactions, Phys. Rev. A 95, 033610 (2017).

\bibitem{Yuan3}  J. B. Yuan and L. M. Kuang, Quantum-discord amplification induced by a quantum phase transition via a cavity-Bose-Einstein-condensate system, Phys. Rev. A 87, 024101  (2013).

\bibitem{Tan1}  Q. S. Tan, Q. T. Xie, and L. M. Kuang, Effects of dipolar interactions on the sensitivity of nonlinear spinor-BEC interterometry, Sci. Rep. 8, 3218 (2018).

\bibitem{Tan2}  Q. S. Tan, J. B. Yuan, G. R. Jin, and L. M. Kuang, Near-Heisenberg-limited parameter estimation precision by a dipolar-Bose-gas reservoir engineering, Phys. Rev. A 96, 063614 (2017).

\bibitem{57} H. T. Ng, and P. T. Leung, Two-mode entanglement in two-component Bose-Einstein condensates, Phys. Rev. A 71, 013601 (2005).

\bibitem{58} H. T. Ng, and K. Burnett, Entanglement between atomic condensates in an optical lattice: Effects of interaction range, Phys. Rev. A 75, 023601 (2007).

\bibitem{55} G. J. Milburn, J. Corney, E. M. Wright, and D. F. Walls, Quantum dynamics of an atomic Bose-Einstein condensate in a double-well potential, Phys. Rev. A 55, 4318 (1997).

\bibitem{56} J. A. Dunningham, M. J. Collett, and D. F. Walls, Quantum state of a trapped Bose-Einstein condensate, Phys. Lett. A 245, 49 (1998).

\bibitem{OMandel} O. Mandel, M. Greiner, A. Widera, T. Rom, T. W. H\"{a}nsch, and I. Bloch, Coherent transport of neutral atoms in spin-dependent optical lattice potentials, Phys. Rev. Lett. 91.010407 (2003).

\bibitem{MBruderer} M. Bruderer, and D. Jaksch, Probing BEC phase fluctuations with atomic quantum dots, New Journal of Physics 8. 87 (2006).

\bibitem{60} S. Inoyue, M. R. Andrews, J. Stenger, H. -J. Miesner, D. M. Stamper-Kurn, and W. Ketterle, Observation of Feshbach resonances in a Bose-Einstein condensate, Nature (London) 392, 151 (1998).

\bibitem{61} S. L. Cornish, N. R. Claussen, J. L. Roberts, E. A. Cornell, and C. E. Wieman, Stable $\sideset{^{85}}{}{\mathop{\mathrm{Rb}}}$ Bose-Einstein condensates with widely tunable interactions, Phys. Rev. Lett. 85, 1795 (2000).

\bibitem{62} A. Marte, T. Volz, J. Schuster, S. D\"{u}rr, G. Rempe, E. G. M. van Kempen, and B. J. Verhaar, Feshbach resonances in Rubidium 87: precision measurement and analysis, Phys. Rev. Lett. 89, 283202 (2002).

\bibitem{63} G. Roati, M. Zaccanti, C. D'Errico, J. Catani, M. Modugno, A. Simoni, M. Inguscio, and G. Modugno, $\sideset{^{39}}{}{\mathop{\mathrm{K}}}$ Bose-Einstein condensate with tunable interactions, Phys. Rev. Lett. 99, 010403 (2007).

\bibitem{Titulaer}  U. M. Titulaer, and R.J. Glauber, Density operators for coherent fields, Phys. Rev. 145, 1041 (1966).

\bibitem{Bialynicka}  Z. Bialynicka-Birula, Properties of the generalized coherent state, Phys. Rev. 173, 1207 (1968).

\bibitem{Stoler}  D. Stoler, Generalized coherent states, Phys. Rev. D 4, 2309 (1971).

\bibitem{Kuang1}  L. M. Kuang and L. Zhou, Generation of atom-photon entangled states in atomic Bose-Einstein condensate via electromagnetically induced transparency, Phys. Rev. A 68, 043606 (2003).

\bibitem{Kuang2}  L. M. Kuang, Z.B. Chen, and J. W. Pan, Generation of entangled coherent states for distant Bose-Einstein condensates via electromagnetically induced transparency, Phys. Rev. A 76, 052324 (2007).

\bibitem{YXHuang} Y. X. Huang, Q. S. Tan, L. B. Fu, and X. G. Wang, Coherence dynamics of a two-mode Bose-Einstein condensate coupled with the environment, Phys. Rev. A 88, 063642 (2013).

\bibitem{Opanchuk} B. Opanchuk, L. Rosales-Z\'{a}rate, R. Y. Teh, and M. D. Reid, Quantifying the mesoscopic quantum coherence of approximate NOON states and spin-squeezed two-mode Bose-Einstein condensates, Phys. Rev. A 94, 062125 (2016).

\bibitem{Greiner}  M. Greiner, O. Mandel, T. W. H\"{a}nsch, and I. Bloch, Collapse and revival of the matter wave field of a Bose-Einstein condensate, Nature, 491,51 (2002).

\bibitem{Shore}  B. W. Shore, P. L. Knight, The Jaynes-Cummings Model, J. Mod. Opt. 40, 1195 (1993).

\bibitem{KLM}  L. M. Kuang, X. Chen, G. H. Chen, and M. M. Ge, Jaynes-Cummings model with phase damping, Phys. Rev. A 56, 3139 (1997).
\end{thebibliography}
\end{document}